\documentclass[a4paper,11pt]{article}
\pdfoutput=1 
\usepackage{jheppub} 
\usepackage[utf8x]{inputenc}
\usepackage[T1]{fontenc} 
\newcommand{\be}{\begin{equation}}
	\newcommand{\ee}{\end{equation}}
\newcommand{\bea}{\begin{array}}
	\newcommand{\ea}{\end{array}}
\newcommand{\beqa}{\begin{eqnarray}}
	\newcommand{\eeqa}{\end{eqnarray}}

\usepackage{physics}

\usepackage{bm}
\usepackage{bbm}
\usepackage{multirow}
\usepackage[table,xcdraw]{xcolor}
\usepackage{mathtools}
\usepackage{dsfont}
\usepackage{float}
\usepackage{caption}
\usepackage{subcaption}

\usepackage[symbol]{footmisc}

\newcommand{\etal}{\text{et al.}}

\title{\boldmath Entanglement dynamics of coupled oscillators
		from Gaussian states}

\author[a]{Cemal Dinç,}
\author{Onur Oktay}

\emailAdd{cedinc@metu.edu.tr}
\emailAdd{oktay24005@gmail.com}

\affiliation[a]{Middle East Technical University, Department of Physics,\\Dumlupinar Boulevard, 06800, Ankara, Turkey}

\abstract{In this work, we explore the dynamics of entanglement of an isolated
	 quantum 
	 system consisting of two time-dependent, coupled harmonic oscillators. Through the use of a numerical method that relies on the estimation of the system's Wigner representation by a specific Gaussian function, we investigate the time evolution of the entanglement entropy after an instant quench in the inherent parameters of the system.
	 Besides, from the comparison of the
	 results obtained from the analytical expression for the time-dependent von Neumann entropy with the numerically computed entropy data,
	  the effectiveness of the numerical method is tested for a variety of angular frequency combinations. Also, we analyze how the entropy of entanglement change as a function of time.}  

\makeatletter
\gdef\@fpheader{}
\makeatother

\notoc  

\begin{document} 
	
	\maketitle
	\flushbottom
	
	\newpage
	
\section{Introduction}

Entanglement is one of the most distinctive features of the quantum mechanics. The counter-intuitive behavior of quantum entanglement, which produces various paradoxical physical phenomena, has confounded the physics community for a very long time. Yet, quantum entanglement has gained a revived interest with the developments in quantum information processing, as it plays a indispensable role, 
being the source of quantum computation, quantum communication \cite{adesso2014continuous,braunstein2005quantum,horodecki2009quantum,nielsen2002quantum,weedbrook2012gaussian}, quantum cryptography \cite{ekert1991quantum} and metrology \cite{nichols2018multiparameter}.

Regarding the entanglement between subsystems constituting a composite system, over the last couple of decades, there have been several attempts to quantify the degree of entanglement in a given bipartite quantum state. One of the measures proposed is the entropy of entanglement or entanglement entropy, which is a natural measure of the degree of quantum correlations for pure  bipartite states in composite systems. The first motivations for studying the features of entanglement entropy came from the physics of black holes. In the seminal work of Sorkin 
\etal\hspace{0.5 mm}\cite{Bombelli:1986rw}, the analytical expression for the ground state entanglement entropy of a system of coupled harmonic oscillators has been derived. Following this milestone, the analysis of \cite{Bombelli:1986rw} has been extended to the bosonic scalar fields in reference \cite{Srednicki:1993im}. In his pioneering work pointing out a possible connection between black hole thermodynamics and entanglement entropy, Srednicki has derived an area law, which scales as the boundary surface dividing the two subsytems and shows a high similarity to the intrinsic entropy formula of a black hole.

On the other hand, since its widespread adoption, there has been a robust interest from the quantum information community to gain enhanced understanding about entanglement entropy due to its capability of providing accurate descriptions of various physical phenomena related to quantum many-body physics \cite{osborne2002entanglement,osterloh2002scaling,vidal2003entanglement}. Traditionally, the study of entanglement in the scope of quantum information theory
has primarily concerned itself with the research in 
discrete physical systems with quantum states defined in finite dimensional Hilbert spaces. Nevertheless, although a daunting complexity arises due to the structure of entangled states in infinite dimensional Hilbert spaces, there has been a growing curiosity for investigating entanglement in continuous variable systems. In this regard, it is essential to remark that 
a particular family of continuous states called the Gaussian states, which are   quantum states that are represented by Gaussian-Wigner quasiprobability functions, have provided the means to explore the dynamics of entanglement in continuous variable systems \cite{Schumaker1986317}. These states naturally arise as the 
thermal or ground states of any bosonic physical system described by at most second order quadrature Hamiltonians in canonical coordinates. Despite the enormous complexity of the underlying infinite dimensional Hilbert space, they can be completely characterized by finite number of degrees of freedom. Besides, Gaussian states play an important role in quantum optics, as they can be easily created and manipulated in a vast number of optical setups. Furthermore, they naturally appear in areas of physics like ion traps, gases of cold atoms and optomechanics \cite{adesso2014continuous,olivares2012quantum}. 

In this paper, our main interest is to analyze the dynamics of entanglement for
an isolated quantum system consisting of two coupled harmonic oscillators after a sudden quench. The departure point of our analysis is the use of a numerical method (the so-called Gaussian state approximation) that relies on the estimation of the system's Wigner representation by a Gaussian function that is characterized by time-dependent wave packet centers and dispersion \cite{Buividovich:2017kfk,Buividovich:2018scl}. 
The paper is structured as follows. Section \ref{sec1} starts out
with brief introductions to the system of two coupled oscillators and the formalism of covariance matrices, which is followed by the derivation of the initial covariance matrix that is employed in the simulations. In section \ref{secEE}, we first describe how the covariance matrix representations of Gaussian states can be utilized to numerically compute entanglement entropy.
Subsequently, the derivation of the analytical expression for the time-dependent von Neumann entropy, which is originated from the exact solutions of the time-dependent Schrödinger equation, is reviewed. In section \ref{secNumr}, we   
investigate the time evolution of the bipartite entanglement entropy within the Gaussian state approximation by utilizing the covariance matrix representations of Gaussian states and compare the numerically obtained results with the findings from  the analytical entropy expression. Lastly, section \ref{Concs} is devoted to
conclusions and outlook.       

\section{Two coupled oscillators and covariance matrix formalism} \label{sec1}
	
We initiate the developments with a consideration of the traditional system of two bosonic oscillators that are coupled by a quadratic Hamiltonian of the form\footnote[2]{In this paper, we work in natural units by setting $\hbar = 1$.}  
\be
\label{HamOsc}
	H = \frac{1}{2} \Big( p_ 1^{\hspace{0.2 mm}2} + p_ 2^{\hspace{0.2 mm}2} \Big) + \frac{1}{2} \Big( \kappa_0(t) \big(x_1^{\hspace{0.2 mm}2} + x_ 2^{\hspace{0.2 mm}2} \big) + \kappa_1(t) {\big({x_ 1} - {x_ 2}\big)}^2 \Big) \,, 
\ee
where $\kappa_0(t)$ and $\kappa_1(t)$ are arbitrary time-dependent parameters. Let us collect the quadrature pairs 
$\big(x_\alpha,p_\beta\big)$ $(\alpha,\beta = 1,2)$ into the vector operator $\boldsymbol{v} \coloneqq \big( x_ 1,x_ 2,p_ 1,p_ 2 \big)^T$, which enables us to express the canonical commutation relations satisfied by these operators, namely $\lbrack x_\alpha , p_\beta \rbrack = i \delta_{\alpha \beta}$, in a more compact form as follows
\be
	\lbrack v_\mu , v_\nu \rbrack = i \Lambda_{\mu \nu} \,, \qquad 
	\Lambda = 
	\begin{pmatrix}
		0_2 & \mathds{1}_2  \\
		-\mathds{1}_2 & 0_2  
	\end{pmatrix},
\ee
where $\mu,\nu = 1,2,3,4$ and the blocks $0_2$ and $\mathds{1}_2$ are the two-dimensional zero and identity matrices. $\Lambda_{\mu \nu}$ are the elements of the antisymmetric matrix $\Lambda$ that is known as the  symplectic form. Let us also note that the matrix transformation $\Lambda$ is orthogonal, i.e. $\Lambda^T \Lambda=\mathds{1}_4$ \cite{Serafini}.   

Having now listed some of the essential features of (\ref{HamOsc}), 
we move on to introduce the Gaussian states. A Gaussian state is defined as any quantum state whose Wigner representation\footnote[3]{The Wigner representation (or function) is the quantum analog of a Liouville probability density. It is a  quasiprobability distribution in the sense that it can take on negative values \cite{peres2006quantum}.} 
is a Gaussian function. Only two parameters, the first and second statistical moments, are required to completely determine a Wigner function. While the first moments are given by the expectation values $\bar{v}_\mu \coloneqq \expval{v_\mu}$, the second moments are defined as \cite{weedbrook2012gaussian,olivares2012quantum,adesso2014continuous}
\be
\label{CovMtx}
	C_{\mu \nu} = {\expval{v_\mu v_\nu}}_{\! c} 
	\coloneqq 
	\frac{1}{2} \expval{\acomm{\delta v_\mu}{\delta v_\nu}} =
	\frac{1}{2}(\expval{\acomm{v_\mu}{v_\nu}}
	-2 \hspace{0.2mm} \bar{v}_\mu \bar{v}_\nu ) 
	\,,
\ee
where $\delta v_\mu \coloneqq v_\mu - \bar{v}_\mu $ are the fluctuation operators
and $C_{\mu \nu}$ are the elements of the real, symmetric matrix $C$, which is referred to as the covariance matrix. The Wigner function can be written by
\be
	\label{Wgnr}
	W(v) = \frac{1}{4 \pi^2 \sqrt{\det(C)}} \exp(-\frac{1}{2} (v-\bar{v})^T C^{-1} (v-\bar{v})) \,,
\ee
where $v \in \mathbb{R}^4$. 
	
To proceed to further, it is convenient to apply the Heisenberg equation of motion to the system of two coupled harmonic oscillators, which results in
\begin{subequations}
		\label{Heis}
		\begin{align}
			\dv{p_ 1}{t} &= -(\kappa_0+\kappa_1)x_ 1 + \kappa_1 x_2\,, \\ 
			\dv{p_ 2}{t} &= -(\kappa_0+\kappa_1)x_ 2 + \kappa_1 x_1 \,, \\ 
			\dv{x_ 1}{t} &= p_ 1 \,, \qquad
			\dv{x_ 2}{t} = p_ 2 \,. 
		\end{align}
\end{subequations} 
The time derivatives of the second moments can be revealed by 
combining the averages of (\ref{Heis}) 
with the averages of  
the operator products of the form $v_\mu v_\nu$ that are both taken over the Gaussian state characterized by (\ref{Wgnr}) \cite{Buividovich:2017kfk}.	
To illustrate, one may 
consider the elements of the first row of the covariance matrix 
$C$.
After a straightforward calculation, the explicit time derivatives of $C_{1 \nu}$ elements are evaluated and shown below 
\begin{subequations}
	\label{GssEqsRw1}
	\begin{align}
		\dv{{\expval{x_1 p_1}}_{\! c}}{t} &= \kappa_1 {\expval{x_1 x_2}}_{\! c}  -(\kappa_0+\kappa_1){\expval{x_1 x_1}}_{\! c}   
		\,, \\ 
		\dv{{\expval{x_1 p_2}}_{\! c}}{t} &= {\expval{p_1 p_2}}_{\! c}  -(\kappa_0+\kappa_1){\expval{x_1 x_2}}_{\! c}  +
		\kappa_1 {\expval{x_1 x_1}}_{\! c} 
		\,, \\ 
		\dv{{\expval{x_1 x_2}}_{\! c}}{t} &= {\expval{x_1 p_2}}_{\! c}  +{\expval{x_2 p_1}}_{\! c}   
		\,, \\
		\dv{{\expval{x_1 x_1}}_{\! c}}{t} &= 2{\expval{x_1 p_1}}_{\! c}  
		\,. 
		\end{align}
\end{subequations} 
	Besides, the remaining six equations describing
	 the time derivatives of the covariance matrix are listed in Appendix \ref{AppEOM}.
	
	In order to study the time evolution of the covariance matrix numerically, apart from determining how $C$ varies in time,
	we also need to specify an initial configuration. 
	With the aim of preparing such a configuration, we start by defining 
	$S \in Sp(4,\mathbb{R})$
	\be
	S = W \oplus W \,, \qquad 
	W = \frac{1}{\sqrt{2}} 
	\begin{pmatrix}
		1 & 1  \\
		-1 & 1  
	\end{pmatrix},
	\ee
	such that $S^T \! \Lambda \hspace{0.2 mm} S = S \hspace{0.2 mm} \Lambda \hspace{0.2 mm} S^T = \Lambda$. Equation
	(\ref{HamOsc}) can be first cast into the matrix form, i.e. 
	\be
	\label{modH1}
	H = \boldsymbol{v}^T H_m \boldsymbol{v} = \frac{1}{2} \boldsymbol{v}^T (H_1 \oplus \mathds{1}_2) \boldsymbol{v}  \,,
	\ee
	and then by diagonalizing $H_m$, (\ref{modH1}) can be rewritten as follows  
	\be
	H = \boldsymbol{v_1}^{\,T} H_d \hspace{0.1mm} \boldsymbol{v_1} \,, \qquad 
	H_d = S \hspace{0.2 mm} H_m \hspace{0.2 mm} S^T = \frac{1}{2} diag\big(\kappa_0,\kappa_0+2\kappa_1,1,1 \big) \,,
	\ee
	where 
	\be
	{\boldsymbol{v_1}} = S \hspace{0.2mm}{\boldsymbol{v}}\,, \qquad 
	H_1 = \begin{pmatrix}
		\kappa_0+\kappa_1 & -\kappa_1  \\
		-\kappa_1 & \kappa_0+\kappa_1  
	\end{pmatrix}.
	\ee
	Now, let us express $H_d$ as a direct sum of diagonal matrices 
	\be
	H_d = H_a \oplus H_b = \frac{1}{2} \begin{pmatrix}
		\kappa_0 & 0  \\
		0 & \kappa_0+2 \kappa_1  
	\end{pmatrix}
	+
	\frac{1}{2}  \mathds{1}_2 \,.
	\ee 
Since $H$ is a quadratic Hamiltonian, it is possible to deduce the ground state covariance matrix from the block diagonal form of $H_d$ as illustrated below \cite{schuch2006quantum}
\begin{align}
		\label{Ci}
		C_i &= \frac{1}{2} S^{-1} \Bigg[ \bigg({H_a}^{-\frac{1}{2}} \sqrt{{H_a}^{\frac{1}{2}} H_b {H_a}^{\frac{1}{2}}} {H_a}^{-\frac{1}{2}} \bigg) \oplus
		\bigg({H_a}^{-\frac{1}{2}} \sqrt{{H_a}^{\frac{1}{2}} H_b {H_a}^{\frac{1}{2}}} {H_a}^{-\frac{1}{2}} \bigg)^{-1}\Bigg] S \nonumber \\
		&= \frac{1}{2} S^{-1} diag\Big(\kappa_0^{-1/2},(\kappa_0+2 \kappa_1)^{-1/2},\kappa_0^{1/2},(\kappa_0+2 \kappa_1)^{1/2}\Big) S
		\nonumber \\ 
		&= \begin{pmatrix}
			\chi_{+} & \chi_{-}  \\
			\chi_{-} & \chi_{+}  
		\end{pmatrix} 
		\oplus
		\begin{pmatrix}
			\xi_{+} & \xi_{-}  \\
			\xi_{-} & \xi_{+} 
		\end{pmatrix} ,
\end{align}
where
\be
	\label{kiZt}
	\chi_\pm = \frac{1}{4}\bigg(\frac{1}{\sqrt{\kappa_0}} \pm \frac{1}{\sqrt{\kappa_0+2 \kappa_1}}\bigg)\,, \qquad 
	\xi_\pm = \frac{1}{4}\bigg(\sqrt{\kappa_0} \pm \sqrt{\kappa_0+2 \kappa_1}\bigg)\,.
\ee
In the next section, a method for computing entanglement entropy from the covariance matrix representation of a Gaussian state is presented. As it will be discussed shortly, the significance of $C_i$ relies on the fact that it corresponds to a pure Gaussian state.
	
\section{Entropy of Entanglement} \label{secEE}
	
The entropy of entanglement is the natural measure of the degree of quantum correlations for pure  bipartite states in composite systems. Let $\rho$ denote the density matrix of the quantum system $\mathcal{S}$ with Hilbert space $\mathcal{H}=\mathcal{H}_a \otimes \mathcal{H}_b$. The entanglement entropy of 
$\mathcal{S}$ can be determined by computing the von Neumann entropy of the reduced density matrix $\rho_a$, which can be obtained by taking the partial trace of $\rho$ over the subspace $\mathcal{H}_b$.
	
\subsection{Entanglement entropy of Gaussian states}
	
Besides the method summarized above, an alternative route of systematically determining the entanglement entropy of a bipartite system is provided by the 
covariance matrix representations of Gaussian states. To get started, we first note that, for $N$ spatial dimensions, the symplectic form and the matrix $\Gamma$ may be introduced as 
\be
	\Gamma = i \Lambda_N C_g \,, \qquad
	\Lambda_N = \begin{pmatrix}
		0_N & \mathds{1}_N  \\
		-\mathds{1}_N & 0_N  
	\end{pmatrix},
\ee
where $C_g$ is the covariance matrix characterizing a generic Gaussian state. 
The symplectic spectrum of $C_g$ can be found
by computing $\abs{eig(\Gamma)}$\footnote[4]{Here, $eig(\Gamma)$ stands for the eigenvalues of $\Gamma$.}, which will yield $N$ pairs of symplectic eigenvalues \cite{serafini2006multimode,adesso2014continuous}. Let $\{e_\tau\}$ for $\tau =1,\dots,N$ denote the symplectic eigenvalues. The uncertainty principle dictates that $e_\tau \geqslant 1/2 \,,$ where the equality only holds for the covariance matrices describing pure Gaussian states.  

As a concrete example, we may focus on $C_i$. From (\ref{Ci}), we have
\be
	\widetilde{\Gamma} = i \Lambda C_i
	= \begin{pmatrix}
		0 & 0 & i\xi_{+} & i\xi_{-} \\
		0 & 0 & i\xi_{-} & i\xi_{+} \\
		-i\chi_{+} & -i\chi_{-} & 0 & 0 \\
		-i\chi_{-} & -i\chi_{+} & 0 & 0 
	\end{pmatrix},
\ee
with characteristic equation
\be
	\lambda^4 - 2 \big(\chi_{+} \xi_{+}+\chi_{-} \xi_{-} \big) \lambda^2 
	+ \big(\chi_{+}^2-\chi_{-}^2\big) \big(\xi_{+}^2-\xi_{-}^2\big)
	= 0 \,.
\ee
Inserting (\ref{kiZt}) into the solution set of $\lambda$ gives 
\be
	\abs{eig \big( \widetilde{\Gamma} \big) } = (1/2,1/2,1/2,1/2) \,,
\ee
which allows us to infer that the symplectic eigenvalues of $C_i$ should be equal to $(1/2,1/2)$. Therefore, it is safe to conclude that $C_i$ characterizes a pure Gaussian state.
	
In order to quantify the bipartite entanglement between the subsystems constituting (\ref{HamOsc}), we need specify the covariance matrix analog of a reduced density matrix. In this regard, it is useful to remark that taking a partial trace is a straightforward task in the covariance matrix formalism.
To elaborate, for the case of two coupled harmonic oscillators, the reduced covariance matrix representing the first subsystem, namely $C_1$, can be derived by extracting the rows and columns corresponding to the first oscillator from the covariance matrix of the whole system $C$ that is previously defined in (\ref{CovMtx}). Written in explicit form, we have
\be
	C_1 = \begin{pmatrix}
		{\expval{x_1 \hspace{0.2mm} x_1}}_{\! c}  & {\expval{x_1 \hspace{0.2mm} p_1}}_{\! c}   \\
		{\expval{p_1 \hspace{0.2mm} x_1}}_{\! c}  & {\expval{p_1 \hspace{0.2mm} p_1}}_{\! c}   
	\end{pmatrix}.
\ee
The von Neumann entropy of a Gaussian state described by $C_1$ is defined as
\be
	\label{S_gss}
	S_c =  \bigg( d+\frac{1}{2} \bigg) \ln(d+\frac{1}{2} ) 
	-  \bigg( d-\frac{1}{2} \bigg) \ln(d-\frac{1}{2} ) \,,
\ee 
where $d$ is the symplectic eigenvalue of $C_1$ that can found by evaluating $\abs{eig(\Gamma_1)}$ with
\be
	\Gamma_1 = i \Lambda_1 C_1 \,, \qquad
	\Lambda_1 = \begin{pmatrix}
		0 & 1  \\
		-1 & 0  
\end{pmatrix} .
\ee
	
\subsection{Time-dependent wave function and entanglement entropy}
		
On the other hand, the ground state entanglement entropy of the system of two coupled harmonic oscillators can be derived analytically for the time-independent case \cite{Bombelli:1986rw,Srednicki:1993im,Terashima:1999vw}. In order to demonstrate this, let us first introduce the unitary transformation 
\be
	U = W^T = \frac{1}{\sqrt{2}} 
	\begin{pmatrix}
		1 & -1  \\
		1 & 1  
	\end{pmatrix} ,
\ee
which makes it possible to rewrite (\ref{HamOsc}) in the familiar uncoupled form of
\be
	H = \frac{1}{2} \Big(p_\sigma^{\hspace{0.2 mm}2} + p_\varrho^{\hspace{0.2 mm}2} \Big) +
	\frac{1}{2} \Big(\omega_{m0}^{\,2} \hspace{0.2mm} x_\varrho^{\hspace{0.2 mm}2} + \omega_{p0}^{\,2} \hspace{0.2mm} x_\sigma^{\hspace{0.2 mm}2} \Big) \,,
\ee
where the eigenfrequencies are 
\be
	\omega_{m0}=\sqrt{\kappa_0(0)+2 \kappa_1(0)}\,, \qquad \omega_{p0}=\sqrt{\kappa_0(0)}\,.
\ee
The original and uncoupled coordinates are related by
\be
   \big(p_\varrho,p_\sigma \big)^T = U
	\hspace{0.2mm} \boldsymbol{p_v} \,, 
	\qquad
	\big(x_\varrho,x_\sigma \big)^T = U
	\hspace{0.2mm} \boldsymbol{x_v} \,, 
\ee 
where $\boldsymbol{p_v} \coloneqq \big(p_ 1,p_ 2 \big)^T$ and  $\boldsymbol{x_v} \coloneqq \big(x_ 1,x_ 2 \big)^T$. The ground state wave function can be expressed in terms of the eigenmodes and eigenfrequencies as shown below
\be
	\label{IniWvf}
	\psi_0 \big( x_\varrho,x_\sigma \big) = \mathcal{N} \exp(-\frac{1}{2} \Big(\omega_{m0} \hspace{0.2 mm} x_\varrho^{\hspace{0.2 mm}2} + 
	\omega_{p0} \hspace{0.2 mm} x_\sigma^{\hspace{0.2 mm}2} \Big) ),
\ee
where $\mathcal{N}= {\big( \omega_{m0} \hspace{0.2mm} \omega_{p0}/ \pi^2 \big)}^{1/4} $ is the normalization factor. The reduced density matrix $\rho_1$ 
can be obtained by first forming the density matrix describing the entire system
in the original coordinates and then tracing over the second oscillator. In the integral form, we have
\be
	\label{rho1}
	\rho_1 = \int_{-\infty}^{\infty}  \psi_0(x_1,x_2) 
	{\big[ \psi_0\big(x^\prime_1,x_2\big)\big]}^*  \dd{x_2}  .
\ee
Upon evaluating (\ref{rho1}), $\rho_1$ may be written by
\be
	\rho_1\big(x_1,x_1^\prime \big) =
	\sqrt{\frac{2}{\pi}
		\bigg( \frac{\omega_{m0} \hspace{0.2mm} \omega_{p0}}{\omega_{m0}+\omega_{p0}} \bigg)} 
 	\hspace{0.2 mm}
	\exp{ x_1 x_1^\prime \vartheta_{10} - \frac{x_1^{\hspace{0.2 mm}2} + {x_ 1^\prime}^{\hspace{0.2 mm}2}}{2} 
		\hspace{0.2mm}	\vartheta_{20}} \,.
\ee
The frequency dependence of $\vartheta_{10}$ and $\vartheta_{20}$ can be detailed as
\be
	\vartheta_{10} = \frac{1}{4}  \frac{\big(\omega_{m0}-\omega_{p0}\big)^2}{\omega_{m0}+\omega_{p0}} \,, \qquad
	\vartheta_{20} = \frac{\omega_{m0}+\omega_{p0}}{4}+\frac{\omega_{m0} \hspace{0.2mm} \omega_{p0}}{\omega_{m0}+\omega_{p0}} \,.
	\ee
After a thorough investigation of the eigenvalue equation, the eigenvalues of $\rho_1$ can be found out and expressed in terms of $\vartheta_\alpha$ as follows
\be
	p_k =\zeta^k (1-\zeta) \,, \qquad 
	\zeta = \frac{\vartheta_{10}}{\vartheta_{20}+
		\sqrt{\vartheta_{20}^{\hspace{0.4 mm}2}-\vartheta_{10}^{\hspace{0.4 mm}2}}}
	\,,
\ee
which may be fed into the well-known entropy formula of
\be
	S = - \sum_{k=0}^{\infty} p_k \log(p_k) \,,
\ee
resulting in
\be
	\label{S_noTime}
	S(\zeta) = \frac{\zeta}{1-\zeta} \log(\frac{1}{\zeta}) + \log(\frac{1}{1-\zeta}) \,. 
\ee
	
Recently, there has been growing interest in exploring the dynamics of entanglement in a system of $N$ coupled oscillators after a quantum quench \cite{ghosh2018entanglement,park2018dynamics,park2019dynamics}. To demonstrate how (\ref{S_noTime}) vary in time, we may follow reference \cite{ghosh2018entanglement} and consider the time-dependent Schrödinger equation in uncoupled coordinates defined by 
\be
	\label{UncpSch}
	i \pdv{\psi}{t} = -\frac{1}{2} \Bigg( \pdv[2]{\psi}{x_\varrho} + \pdv[2]{\psi}{x_\sigma} \bigg)
	+
	\frac{1}{2} 
	\Big(\omega_{m}^{\,2} \hspace{0.2mm} x_\varrho^{\hspace{0.2 mm}2} + \omega_{p}^{\,2} \hspace{0.2mm} x_\sigma^{\hspace{0.2 mm}2} \Big) \psi \,,
\ee
where $\psi = \psi \big(x_\varrho,x_\sigma,t \big) $ and the eigenfrequencies are given by
\be
	\omega_{m}(t)=\sqrt{\kappa_0(t)+2 \kappa_1(t)}\,, \qquad \omega_{p}(t)=\sqrt{\kappa_0(t)}\,.
\ee
Let $E_m$ and $E_p$ denote the energies of the uncoupled systems at the initial time $t=0$. By choosing (\ref{IniWvf}) as the initial wave function and subsequently solving (\ref{UncpSch}), we arrive at the time-dependent wave function:
\be
	\label{tmPhi}
	\psi \big(x_\varrho,x_\sigma,t \big)
	= 
	\exp(
	\frac{i}{2} \Big(\Delta_1 x_\sigma^{\hspace{0.2 mm}2}
	+ \Delta_2 x_\varrho^{\hspace{0.2 mm}2} \Big)
	-i \hspace{0.2 mm} \Xi(t)
	)
	\psi \bigg(\frac{x_\varrho}{\gamma _2},0 \bigg)
	\psi \bigg(\frac{x_\sigma}{\gamma _1},0 \bigg) \,,
\ee
where 
\be
	\Xi(t) =  E_m \Theta_m(t) +  E_p \Theta_p(t) \,, \qquad 
	\Delta_\alpha= \frac{\dot{\gamma }_\alpha}{ \gamma _\alpha} \,,
\ee
with
\be
	\Theta_m(t) = \int_{0}^{t}\frac{\dd{u}}{\gamma _2^{\hspace{0.2 mm}2}(u)} \,, \qquad 
	\Theta_p(t) = \int_{0}^{t}\frac{\dd{u}}{\gamma _1^{\hspace{0.2 mm}2}(u)} \,.
\ee
$\gamma _\alpha(t)$ are governed by the Ermakov equations \cite{lohe2008exact}
\begin{subequations}
		\label{ErmEqs}
		\begin{align}
			\ddot{\gamma}_1 + \omega_{p}^{\hspace{0.2 mm}2}(t) \gamma_1(t) &= \frac{\omega_{p0}^{\hspace{0.2 mm}2}}{\gamma_1^{\hspace{0.2 mm}3}(t)}   \,,
			\\  
			\ddot{\gamma}_2 + \omega_{m}^{\hspace{0.2 mm}2}(t) \gamma_2(t) &= 
			\frac{\omega_{m0}^{\hspace{0.2 mm}2}}{\gamma_2^{\hspace{0.2 mm}3}(t)}  \,,
		\end{align}
\end{subequations} 
with the conditions $\gamma_\alpha(0)=0$ and $\dot{\gamma }_\alpha(0)=0$\footnote[5]{These conditions are specifically chosen to ensure unitarity \cite{lohe2008exact} and will remain fixed for the rest of this work.}. Repeating the procedure detailed earlier\footnote[6]{See equations (\ref{rho1}) to (\ref{S_noTime}).}, the time evolution of the bipartite entanglement can be determined as
\be
	\label{Sanlytc}
	S_a(t) =  \frac{\Pi}{1-\Pi} \log(\frac{1}{\Pi}) + \log(\frac{1}{1-\Pi}) \,, 
\ee
where
\be 
	\label{BgPi}
	\Pi(t) = \frac{\vartheta_{1}}{\vartheta_{2}+
		\sqrt{\vartheta_{2}^{\hspace{0.2 mm}2}-\vartheta_{1}^{\hspace{0.2 mm}2}}}
	\,.
\ee
In (\ref{BgPi}), $\vartheta_{1}$ and $\vartheta_{2}$ are time-dependent functions given by
\be
	\vartheta_{1} = \frac{\wp_{-}^{\hspace{0.2 mm}2} +  (\Delta_1- \Delta_2)^2 }{4 \wp_{+}} \,, \qquad 
	\vartheta_{2} = \frac{\wp_{+}}{2} -
	\frac{\wp_{-}^{\hspace{0.2 mm}2}
		- (\Delta_1- \Delta_2)^2 }{4 \wp_{+}}
	\,. 
	\ee
with
\be
	\label{EqUps}
	\wp_\pm = \frac{\omega_{p0}}{\gamma_1^{\hspace{0.2 mm}2}} \pm \frac{\omega_{m0}}{\gamma_2^{\hspace{0.2 mm}2}}  \,.
\ee

\subsubsection{Quenched model}
	
The set of nonlinear equations listed in (\ref{ErmEqs}) can be studied numerically with the help of an iterative algorithm implemented in computer code.
Alternatively, if a quenched model is assumed, explicit analytical solutions of $\gamma _\alpha(t)$ can be written out. 
Suppose that the frequencies $\omega_{p}$ and $\omega_{m}$ are instantly quenched after the initial time. Then, explicitly, we have
\be
	\label{QnchMec}
	\omega_{p}(t) =
	\begin{dcases}
		\omega_{p0} \,, & t = 0 \\
		\omega_{p1} \,, & t > 0
	\end{dcases} \,,
	\qquad
	\omega_{m}(t) =
	\begin{dcases}
		\omega_{m0}, & t = 0 \\
		\omega_{m1}, & t > 0
	\end{dcases} \,.
\ee
Under this restriction, the solution set of (\ref{ErmEqs}) becomes
\begin{subequations}
	\begin{align}
			\gamma_1(t) &= {(\varepsilon_{-} \cos(2 \hspace{0.3mm} \omega_{p1} \hspace{0.2mm} t) + \varepsilon_{+})}^{\frac{1}{2}}  \,,
			\\  
			\gamma_2(t) &= {(\epsilon_{-} \cos(2 \hspace{0.3mm} \omega_{m1} \hspace{0.2mm} t) + \epsilon_{+})}^{\frac{1}{2}}  \,,
	\end{align}
\end{subequations} 
from which it is easy to see that $\Delta_\alpha$ are now given by
\be
	\Delta_1 = -\frac{ \omega_{p1} \hspace{0.2 mm} \varepsilon_{-} \sin(2 \hspace{0.3mm} \omega_{p1} \hspace{0.2mm} t) }{\gamma_1^{\hspace{0.2 mm}2}} \,, 
	\qquad
	\Delta_2 = -\frac{\omega_{m1} \hspace{0.2 mm} \epsilon_{-} \sin(2 \hspace{0.3mm} \omega_{m1} \hspace{0.2mm} t)}{\gamma_2^{\hspace{0.2 mm}2}} \,, 
\ee
where
\be
	\varepsilon_{\pm} = \frac{\omega_{p1}^{\hspace{0.2 mm}2} 
		\pm \omega_{p0}^{\hspace{0.2 mm}2}}
	{2 \hspace{0.2 mm} \omega_{p1}^{\hspace{0.2 mm}2}} \,, 
	\qquad
	\epsilon_{\pm} =  \frac{\omega_{m1}^{\hspace{0.2 mm}2} 
		\pm \omega_{m0}^{\hspace{0.2 mm}2}}
	{2 \hspace{0.2 mm} \omega_{m1}^{\hspace{0.2 mm}2}} \,.
\ee
Together with equations (\ref{BgPi}) to (\ref{EqUps}), the last three relations
obviously suffice to determine $S_a(t)$, and hence explore the dynamics of entanglement in the quenched model.
	
\section{Numerical results} \label{secNumr}   
	
This section is devoted to an investigation of the dynamics of entanglement observed in the system of two coupled harmonic oscillators. In order to provide a comprehensive analysis, we carry out numerical simulations 
of the time evolution of (\ref{S_gss}) for various distinct frequency 
settings. After discretizing the equations of motion given in (\ref{GssEqsRw1}) and (\ref{AppEOM}), an iterative algorithm is utilized to
solve the discretized equations numerically. As it will be detailed shortly, we start the simulations with initial conditions that are formed from
the elements of the matrix $C_i$, which is listed in (\ref{Ci}). 
For comparison, aside from the numerical findings obtained from the evaluation of $S_c$, we also consider the bipartite entanglement entropy expression of (\ref{Sanlytc}) and calculate $S_a$ for the frequency combinations that are employed in the simulations of (\ref{S_gss}). 
From the numerical computations of $S_c$ and the comparisons of $S_c$ with $S_a$,
we hope to gain valuable insight both into the dynamics of entanglement and the effectiveness of the Gaussian state approximation. 
		
In the computations, a simulation code that is implemented in Matlab is used. The code is executed with a constant time step of $\Delta t = 0.001$.  Due to truncation of digits, errors are inevitable in numerical calculations. In this regard, although (as it is characterizing a pure Gaussian state) $C_i$ is expected to keep representing pure Gaussian states due to the quadratic structure of $(\ref{HamOsc})$ \cite{Buividovich:2017kfk,Buividovich:2018scl}, the cumulative effect of rounding errors could cause a violation of this preserving map and create mixed states.  However, by constantly monitoring the symplectic eigenvalues during the the trial runs of the simulation, we made sure that no such effect is present. 
		 
Having now introduced the basic features of numerical computations, we move on to the details of starting configurations. The simulations are initiated from the previously determined covariance matrix $C_i$ defined by $(\ref{Ci})$. Namely, at the initial time $t=0$, we have	
\begin{subequations}
		\begin{align}
			{\expval{x_1 x_1}}_{\! c} &= {\expval{x_2 x_2}}_{\! c} = \chi_+ \,, \qquad
		    {\expval{x_1 x_2}}_{\! c} = {\expval{x_2 x_1}}_{\! c} = \chi_- \,,	\\ 
			{\expval{p_1 p_1}}_{\! c} &= {\expval{p_2 p_2}}_{\! c} = \xi_+ \,, \qquad
			{\expval{p_1 p_2}}_{\! c} = {\expval{p_2 p_1}}_{\! c} = \xi_- \,,			
		\end{align}
\end{subequations} 
while the rest of the initial conditions are zero. Before moving on to the presentation of numerical findings, we need to point out that the simulations are started with frequencies $\omega_{p0}$ and $\omega_{m0} \,$. Upon the completion of the first time increment $\Delta t$, these frequencies are updated to $\omega_{p1}$ and $\omega_{m1} \hspace{0.2mm}$, which remain unchanged for the second time increment and the rest of the computations. This is how the quenching mechanism of $(\ref{QnchMec})$ is included in the implementation of the algorithm.
 		  
\begin{figure}[!htb]
 	\centering
    \begin{subfigure}[!htb]{.496\textwidth}
 		  		\centering
 		  		\includegraphics[width= 1\linewidth]{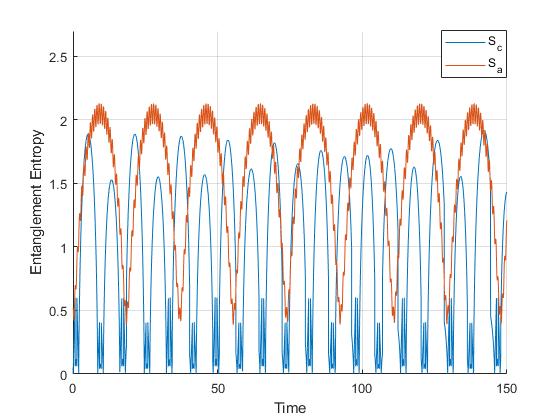}  
 		  		\caption{$\omega_{p1}=0.17 \,, \,\,\, \omega_{m1}=4.17$}
 		  		\label{fig:fige1a}
 	\end{subfigure}	
 	\begin{subfigure}[!htb]{.496\textwidth}
 		  		\centering
 		  		\includegraphics[width=
 		  		1\linewidth]{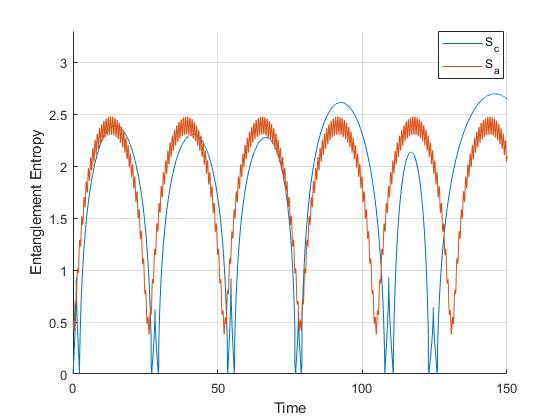}  
 		  		\caption{$\omega_{p1}=0.12 \,, \,\,\, \omega_{m1}=4.12$}
 		  		\label{fig:fige1b}
 	\end{subfigure}	
 	\begin{subfigure}[!htb]{.496\textwidth}
 		  		\centering
 		  		\includegraphics[width=
 		  		1\linewidth]{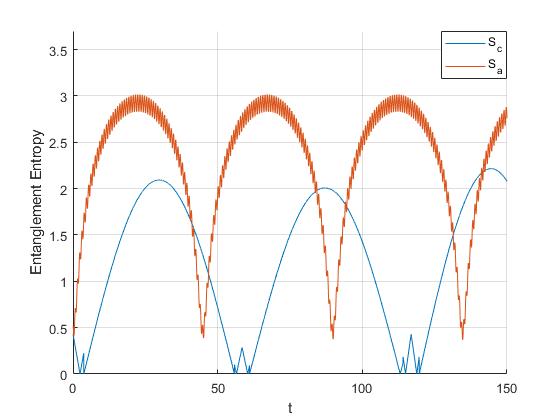}  
 		  		\caption{$\omega_{p1}=0.07 \,, \,\,\, \omega_{m1}=4.07$}
 		  		\label{fig:fige1c}
 	\end{subfigure}	
 	\begin{subfigure}[!htb]{.496\textwidth}
 		  		\centering
 		  		\includegraphics[width=
 		  		1\linewidth]{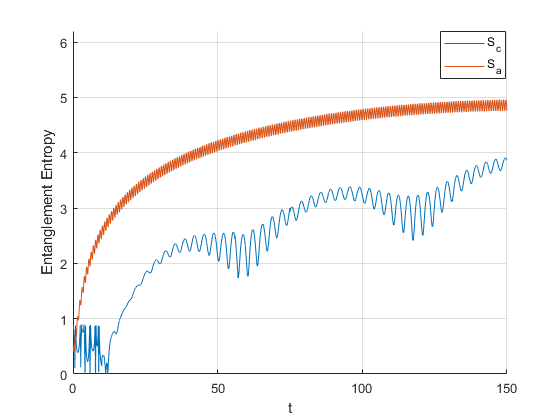}  
 		  		\caption{$\omega_{p1}=0.01 \,, \,\,\, \omega_{m1}=4.01$}
 		  		\label{fig:fige1d}
   \end{subfigure}	
   \caption{The time-dependence of the von Neumann entropies $S_a$ and $S_c$. Here the initial eigenfrequencies are $\omega_{p0}=0.9$ and $\omega_{m0}=4.9$.}
   \label{fig:fige1}
 \end{figure}

The comparison between two distinct measures of von Neumann entropy and the long-time behavior analysis of $S_c \hspace{0.2mm}$,
both of which would shed light on the effectiveness of the Gaussian state approximation, are made through the use of figures and tables.   
In Figure \ref{fig:fige1}, we present sample plots for the time series of $S_a$ and $S_c$ after a sudden quantum quench. While the initial frequencies are held fixed at $\omega_{p0}=0.9$ and $\omega_{m0}=4.9$, the final frequencies, i.e. $\omega_{p1}$ and $\omega_{m1}$, are varied. The first thing we can immediately observe from the plots is that the oscillation frequencies of both $S_a$ and
 $S_c$ decrease with reduced final frequencies. In Figure \ref{fig:fige1a},
 the ratio of the oscillation periods of two entropy measures
 is approximately equal to $2.3\hspace{0.2mm}$. When both $\omega_{p1}$ and $\omega_{m1}$ are reduced by $0.05$, this ratio shows the tendency to 
 converge to $1$ as it can be observed from Figure \ref{fig:fige1b}. Furthermore,
 in the same graph, the difference between $S_a$ and the numerically computed entanglement entropy $S_c$ is the smallest among all the plots.
 In particular, for $t\lessapprox 83\hspace{0.2mm}$, the Gaussian state approximation is highly effective for the frequency combination depicted in Figure \ref{fig:fige1b}. Similarly, in Figures \ref{fig:fige1c} and \ref{fig:fige1d}, despite an apparent increase in the difference of the oscillation amplitudes, $S_c$ certainly shows the ability to follow the oscillatory motion of $S_a$.   
 	     
 Regarding the possible sources of error in the computations of the numerically calculated entropy, it is essential to note that despite its capability in conserving the von Neumann entropy by evolving pure states into pure states, 
 the Gaussian state approximation describes a non-unitary evolution of the density matrices \cite{Buividovich:2018scl}. 
 This is certainly not true for the case of equation (\ref{Sanlytc}), which is emanated from the exact solutions of the time-dependent Schrödinger equation and describes a unitary time evolution.
 
\begin{figure}[!htb]
 	  	\centering
 	  	\begin{subfigure}[!htb]{.496\textwidth}
 	  		\centering		
 	  		\includegraphics[width=1\linewidth]
 	  		{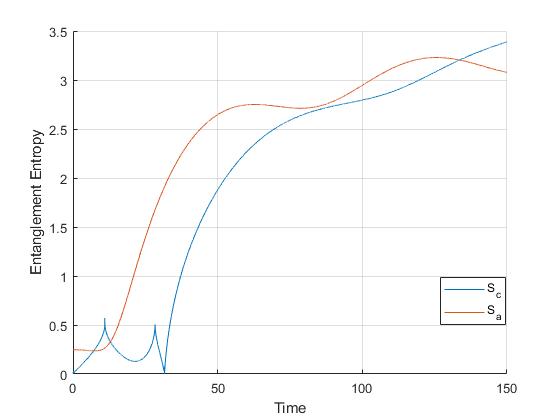}
 	  		\caption{$\omega_{p1}=0.04 \,, \,\,\, \omega_{m1}=0.05$}
 	  		\label{fig:fige2a}
 	  	\end{subfigure}	
 	  	\begin{subfigure}[!htb]{.496\textwidth}
 	  		\centering
 	  		\includegraphics[width=1\linewidth]
 	  		{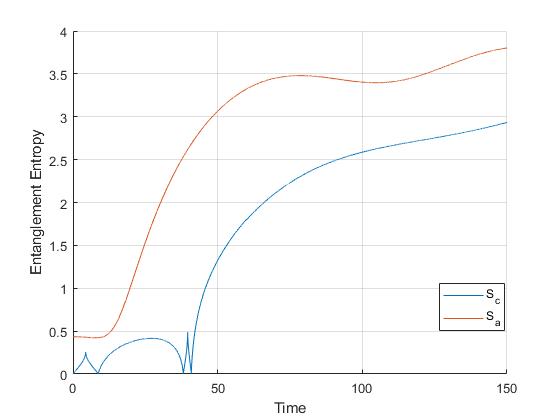}
 	  		\caption{$\omega_{p1}=0.03 \,, \,\,\, \omega_{m1}=0.04$}
 	  		\label{fig:fige2b}
 	  	\end{subfigure}	
 	  	\begin{subfigure}[!htb]{.496\textwidth}
 	  		\centering
 	  		\includegraphics[width=1\linewidth]
 	  		{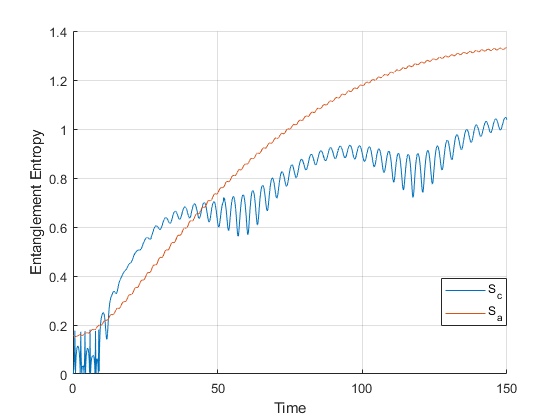}  
 	  		\caption{$\omega_{p1}=1.49 \,, \,\,\, \omega_{m1}=1.5$}
 	  		\label{fig:fige2c}
 	  	\end{subfigure}	
 	  	\begin{subfigure}[!htb]{.496\textwidth}
 	  		\centering
 	  		\includegraphics[width=1\linewidth]
 	  		{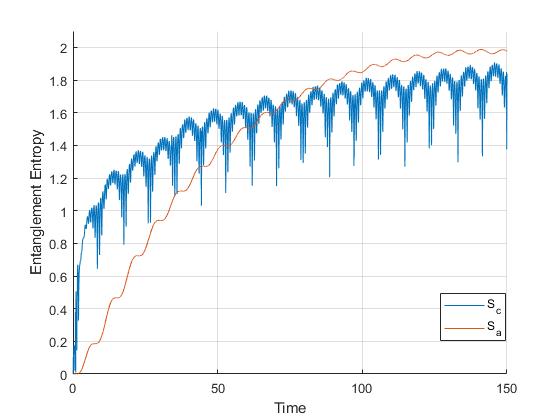}  
 	  		\caption{$\omega_{p1}=0.412 \,, \,\,\, \omega_{m1}=0.423$}
 	  		\label{fig:fige2d}
 	  	\end{subfigure}	
 	  	\caption{Plots of $S_a$ and $S_c$ vs. Time. Here the initial eigenfrequencies are (a) $\omega_{p0}=0.94$, $\omega_{m0}=2.6$ (b) $\omega_{p0}=1.02$, $\omega_{m0}=4.6$ (c) $\omega_{p0}=0.186$, $\omega_{m0}=0.396$ (d)
 	  	$\omega_{p0}=4.33$, $\omega_{m0}=4.49$ .}
 	  	\label{fig:fige2}
 \end{figure}

 On the other hand, in order to take the effects of the starting frequencies into consideration, we
 illustrate in Figure \ref{fig:fige2} the evolutions of the entanglement entropies with time at four different initial frequency combinations. Despite the presence of periodic ripples in Figure \ref{fig:fige2d}, as it can be clearly seen from all four plots, $S_c$ is capable of following $S_a \hspace{0.2mm}$, which shows an upward trend with time.
 	      
\begin{figure}[!htb]
   	\centering
   	\begin{subfigure}[!htb]{.496\textwidth}
   		\centering
   		\includegraphics[width= 1\linewidth]
   		{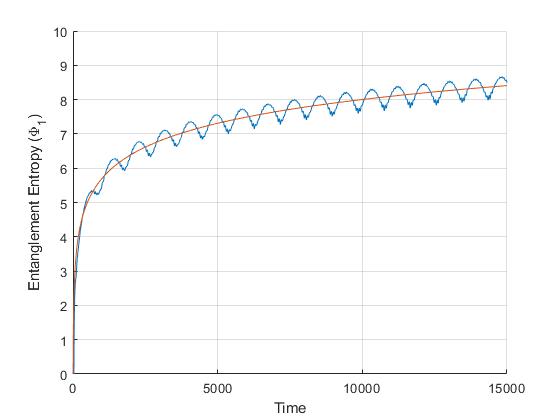}  
   		\caption{$\omega_{p1}=0.04 \,, \,\,\, \omega_{m1}=0.05$}
   		\label{fig:fige3a}
   	\end{subfigure}	
   	\begin{subfigure}[!htb]{.496\textwidth}
   		\centering
   		\includegraphics[width=1\linewidth]
   		{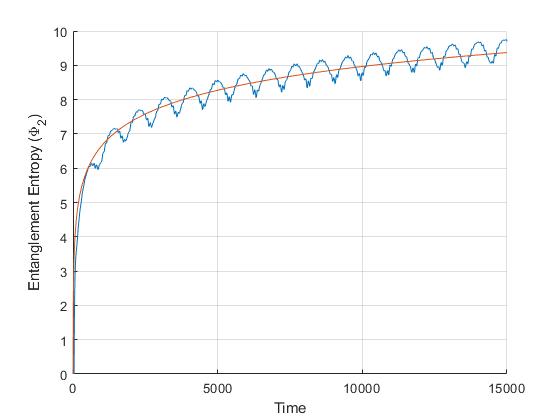}  
   		\caption{$\omega_{p1}=0.03 \,, \,\,\, \omega_{m1}=0.04$}
   		\label{fig:fige3b}
   	\end{subfigure}	
   	\begin{subfigure}[!htb]{.496\textwidth}
   		\centering
   		\includegraphics[width=1\linewidth]
   		{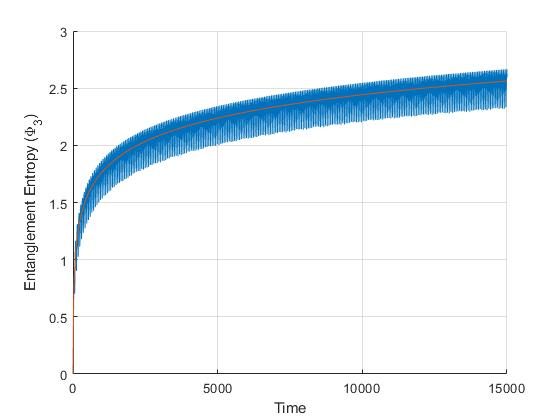}  
   		\caption{$\omega_{p1}=1.49 \,, \,\,\, \omega_{m1}=1.5$}
   		\label{fig:fige3c}
   	\end{subfigure}	
   	\begin{subfigure}[!htb]{.496\textwidth}
   		\centering
   		\includegraphics[width=1\linewidth]
   		{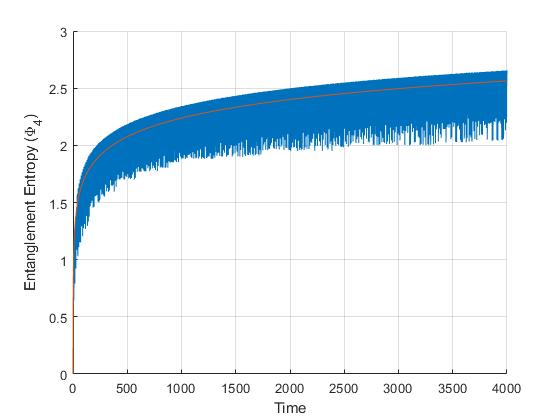}  
   		\caption{$\omega_{p1}=0.412 \,, \,\,\, \omega_{m1}=0.423$}
   		\label{fig:fige3d}
   	\end{subfigure}	
   	\caption{Plots of the von Neumann entropy $S_c$ vs. Time. Here the initial eigenfrequencies are (a) $\omega_{p0}=0.94$, $\omega_{m0}=2.6$ (b) $\omega_{p0}=1.02$, $\omega_{m0}=4.6$ (c) $\omega_{p0}=0.186$, $\omega_{m0}=0.396$ (d)
   		$\omega_{p0}=4.33$, $\omega_{m0}=4.49$.}
   	\label{fig:fige3}
   \end{figure}

Lastly, we would like to report on the results of long-time behavior analysis of $S_c \hspace{0.2mm}$. Figure \ref{fig:fige3} shows the plots of $S_c$ versus time at the frequency combinations used for the preparation of Figure \ref{fig:fige2}. This time we run the code for a sufficient amount of time
   to clearly observe the values that the entanglement entropies converge to. The best-fitting model for the numerical data displayed in Figure  is found to be a logarithmic function in the form
   \be
   \label{logfit}
   \Phi_\mu (t) 
   = u_\mu \log\big( w_\mu t \big) + z_\mu \,.
   \ee
   The fitting parameters of the best fit equations (\ref{logfit}) are listed in Table \ref{table:fitvalues}. The adjusted R-squared values of the fitting curves depicted in Figures \ref{fig:fige3a} - \ref{fig:fige3d} are given by $0.9581$, $0.9394$, $0.9134$ and $0.884$ respectively. Although, due to the 
   apparent increase in the variance of the data illustrated in Figures  \ref{fig:fige3c}-\ref{fig:fige3d}, 
   $\Phi_3$ and $\Phi_4$ fits are not as good in comparison to the
    \begin{table}[H]
   	\centering
   	\begin{tabular}{ | c | c | c | c |}
   		\cline{2-4}
   		\multicolumn{1}{c |}{} &  $u_\mu$ & $w_\mu$ & $z_\mu$ \\ \hline
   		$\Phi_1(t)$ &$1$ &$0.299$ &$0$ \\ \hline
   		$\Phi_2(t)$ &$1$ &$0.783$ &$0$ \\ \hline
   		$\Phi_3(t)$  &$0.295$ & $0.0512$ &$0.604$ \\ \hline
   		$\Phi_4(t)$  &$0.234$ & $14.04$ &$0$ \\ \hline
   	\end{tabular}
   	\caption{$u_\mu \hspace{0.2mm}$, $w_\mu$ and $z_\mu$ values for the fitting curve (\ref{logfit})}
   	\label{table:fitvalues}
   \end{table}
   \noindent fits shown in Figures \ref{fig:fige3a}-\ref{fig:fige3b}, the magnitudes of R-squared values indicates that $\Phi_\mu$ curves still constitute adequate models. Thus, it is safe to conclude that numerically computed von Neumann entropy $S_c$ vary logarithmically with time.   

\section{Conclusions and outlook} \label{Concs}

In this paper, we have considered the dynamics of entanglement of a system of two coupled harmonic oscillators following an instant quantum quench in the inherent parameters of the system. We have performed a detailed numerical analysis of the time evolution of the bipartite entanglement entropy within the Gaussian state approximation, whose accuracy has been verified qualitatively through the comparison with the analytical expression for the time-dependent entanglement entropy. Besides, from the results concerning the change of von Neumann entropy with time, we were able to show through a proper fitting function that entanglement entropy vary logarithmically in time under the same approximation.

Let us also mention a recent development regarding the possible applications of the Gaussian state approximation. Although calculating entanglement entropy in ordinary field theories is a rather difficult task, the numerical computations of the entanglement entropy in the Banks-Fischler-Shenker-Susskind (BFSS) matrix model were recently performed in reference \cite{Buividovich:2018scl} by using the covariance matrix representations of Gaussian states. In this regard, 
a valuable direction of research would be to investigate the time-dependence of the entanglement entropy in a Yang-Mills matrix model with two distinct mass deformation terms, whose emerging chaotic motions and dynamics of thermalization have been recently studied in a series of papers \cite{Baskan:2019qsb,Oktay:2020tzi}. Aside from the mass deformation terms keeping the gauge invariance intact, this model contains the same matrix content as the bosonic part of the BFSS matrix model and provides an ideal system for the use of Gaussian state approximation. Moreover, with the help of the results obtained in \cite{Oktay:2020tzi}, it would be interesting to investigate the possible use of entanglement entropy as a probe of thermalization. Another
challenging direction of development is to generalize the approach
and methods developed in this work for the analysis of the variation of entanglement entropy of two time-dependent, coupled harmonic oscillators to the case of $N$ time-dependent, coupled harmonic oscillators. We hope to report on progress in these directions elsewhere.

\appendix

\section{Time derivatives of $C_{\mu \nu}$}\label{RstGssEOm}

In this appendix, we present equations describing the time derivatives of six distinct second moments. Together with equation (\ref{GssEqsRw1}), these relations define $\partial_t (C_{\mu \nu})$. 
\begin{subequations}
	\label{AppEOM}
	\begin{align}
		\dv{{\expval{p_1 p_1}}_{\! c}}{t} &= 2 \kappa_1 {\expval{x_2 p_1}}_{\! c}  -2 (\kappa_0+\kappa_1){\expval{x_1 p_1}}_{\! c}   
		\,, \\ 
		\dv{{\expval{p_1 x_2}}_{\! c}}{t} &= {\expval{p_1 p_2}}_{\! c} 
		+ \kappa_1 {\expval{x_2 x_2}}_{\! c}  -(\kappa_0+\kappa_1){\expval{x_1 x_2}}_{\! c}  
		\,, \\
		\dv{{\expval{p_1 p_2}}_{\! c}}{t} &= 
		-(\kappa_0+\kappa_1)
		({\expval{x_1 p_2}}_{\! c}  + {\expval{x_2 p_1}}_{\! c})
		+ \kappa_1( {\expval{x_2 p_2}} + {\expval{p_1 x_1}} \\
		&- {\expval{x_2}}{\expval{p_2}} - {\expval{x_1}}{\expval{p_1}}   ) \nonumber
		\,, \\
		\dv{{\expval{x_2 p_2}}_{\! c}}{t} &= -(\kappa_0+\kappa_1){\expval{x_2 	x_2}}_{\! c} + \kappa_1 {\expval{x_1 x_2}}_{\! c} +
		 {\expval{p_2 p_2}}_{\! c}      
		\,, \\ 
		\dv{{\expval{p_2 p_2}}_{\! c}}{t} &= 2\kappa_1{\expval{x_1 p_2}}_{\! c} + -2 (\kappa_0 + \kappa_1) {\expval{x_2 p_2}}_{\! c}       
		\,, \\ 
			\dv{{\expval{x_2 x_2}}_{\! c}}{t} &= 2 {\expval{x_2 p_2}}_{\! c}  
		\,.
   \end{align}
\end{subequations}


\begin{thebibliography}{99}
	
\bibitem{horodecki2009quantum}
R.~Horodecki, P.~Horodecki, M.~Horodecki, and K.~Horodecki,
Rev. Mod. Phys. \textbf{81}, 865 (2009).

\bibitem{braunstein2005quantum}
S.~L.~Braunstein and P.~Van~Loock,
Rev. Mod. Phys. \textbf{77}, 513 (2005).

\bibitem{nielsen2002quantum}
M.~A.~Nielsen and I.~Chuang,
{\it Quantum computation and quantum information}, 
\newblock Cambridge University Press, 2002.	

\bibitem{weedbrook2012gaussian}
C.~Weedbrook, S.~Pirandola, R.~Garc{\'\i}a-Patr{\'o}n,
N.~J.~Cerf, T.~C.~Ralph, J.~H.~Shapiro, and S.~Lloyd,
Rev. Mod. Phys. \textbf{84}, 621 (2012).

\bibitem{adesso2014continuous}
G.~Adesso, S.~Ragy, and A.~R.~Lee,
Open Syst. Inf. Dyn. \textbf{21}, 1440001 (2014).

\bibitem{ekert1991quantum}
A.~K.~Ekert, Phys. Rev. Lett. \textbf{67}, 661 (1991).

\bibitem{nichols2018multiparameter}
R.~Nichols, P.~Liuzzo-Scorpo, P.~A.~Knott, and G.~Adesso,
Phys. Rev. A \textbf{98}, 012114 (2018).

\bibitem{Bombelli:1986rw}
L.~Bombelli, R.~K.~Koul, J.~Lee and R.~D.~Sorkin,
Phys. Rev. D \textbf{34}, 373-383 (1986)
doi:10.1103/PhysRevD.34.373

\bibitem{Srednicki:1993im}
M.~Srednicki,
Phys. Rev. Lett. \textbf{71}, 666-669 (1993)
doi:10.1103/PhysRevLett.71.666
[arXiv:hep-th/9303048 [hep-th]].

\bibitem{osborne2002entanglement}
T.~J.~Osborne and M.~A.~Nielsen,
Phys. Rev. A \textbf{66}, 032110 (2002).

\bibitem{osterloh2002scaling}
A.~Osterloh, L.~Amico, G.~Falci, and R.~Fazio,
Nature \textbf{416}, 608 (2002).

\bibitem{vidal2003entanglement}
G.~Vidal, J.~I.~Latorre, E.~Rico, and A.~Kitaev,
Phys. Rev. Lett. \textbf{90}, 227902 (2003).

\bibitem{Schumaker1986317}
B.~L.~Schumaker,
Phys. Rep. \textbf{135}, 317 (1986).

\bibitem{olivares2012quantum}
S.~Olivares,
Eur. Phys. J. Spec. Top. \textbf{203}, 3 (2012).

\bibitem{Buividovich:2017kfk}
P.~Buividovich, M.~Hanada and A.~Sch\"afer,
EPJ Web Conf. \textbf{175}, 08006 (2018)
doi:10.1051/epjconf/201817508006
[arXiv:1711.05556 [hep-th]].

\bibitem{Buividovich:2018scl}
P.~V.~Buividovich, M.~Hanada and A.~Sch\"afer,
Phys. Rev. D \textbf{99}, no.4, 046011 (2019)
doi:10.1103/PhysRevD.99.046011
[arXiv:1810.03378 [hep-th]].

\bibitem{Serafini} 
A.~Serafini,
{\it Quantum continuous variables: a primer of theoretical methods}, 
\newblock CRC Press, 2017.	

\bibitem{peres2006quantum}
A.~Peres,
{\it Quantum theory: concepts and methods}, 
vol.~57,
\newblock Springer Science \& Business Media, 2006.

\bibitem{schuch2006quantum}
N.~Schuch, J.~I.~Cirac, and M.~M.~Wolf,
Commun. Math. Phys. \textbf{267}, 65 (2006).

\bibitem{serafini2006multimode}
A.~Serafini,
Phys. Rev. Lett. \textbf{96}, 110402 (2006).

\bibitem{Terashima:1999vw}
H.~Terashima,
Phys. Rev. D \textbf{61}, 104016 (2000)
doi:10.1103/PhysRevD.61.104016
[arXiv:gr-qc/9911091 [gr-qc]].

\bibitem{ghosh2018entanglement}
S.~Ghosh, 
K.~S.~Gupta, and S.~C.~L.~Srivastava,
EPL \textbf{120}, 50005 (2018).

\bibitem{park2018dynamics}
D.~Park,
Quantum Inf. Process. \textbf{17}, 147 (2018).

\bibitem{park2019dynamics}
D.~Park,
Quantum Inf. Process. \textbf{18}, 282 (2019).

\bibitem{lohe2008exact}
M.A.~Lohe,
J. Phys. A: Math. Theor. \textbf{42}, 035307 (2008).

\bibitem{Baskan:2019qsb}
K.~Ba\c{s}kan, S.~K\"urk\c{c}\"uoǧlu, O.~Oktay and C.~Ta\c{s}c\i{},
JHEP \textbf{10}, 003 (2020)
doi:10.1007/JHEP10(2020)003
[arXiv:1912.00932 [hep-th]].

\bibitem{Oktay:2020tzi}
O.~Oktay,
[arXiv:2010.12927 [hep-th]].


\end{thebibliography}
\end{document}